\documentclass[lettersize,journal]{IEEEtran}
\usepackage{amsmath,amsfonts}
\usepackage{algorithmic}
\usepackage{algorithm}
\usepackage{array}
\usepackage[caption=false,font=normalsize,labelfont=sf,textfont=sf]{subfig}
\usepackage{textcomp}
\usepackage{stfloats}
\usepackage[hyphens]{url}
\usepackage{verbatim}
\usepackage{cite}

\usepackage{booktabs}
\usepackage{multirow}
\usepackage{graphicx}
\usepackage{amssymb}
\usepackage{acronym}
\usepackage{xspace}
\usepackage[export]{adjustbox}

\usepackage{hyperref}
\usepackage[table]{xcolor}

\usepackage{regexpatch}
\usepackage{todonotes}
\usepackage{adjustbox}
\usepackage{enumitem}
\usepackage{multirow}
\usepackage{tabularx}
\usepackage{tabularray}
\usepackage{pifont} 
\definecolor{darkgreen}{rgb}{0.0, 0.5, 0.0}

\newcommand{\toolname}{\textsc{Dirty-Waters}\xspace}
\newcommand{\toolnameaction}{\textsc{Dirty-Waters-Action}}

\usepackage{nccmath}
\usepackage{xspace}
\usepackage{xargs}
\usepackage{microtype}
\usepackage{tabularx}
\usepackage{booktabs}
\usepackage{listings}
\usepackage{parcolumns}
\usepackage[many]{tcolorbox}
\usepackage{float}
\usepackage{soul}
\usepackage{footmisc}
\usepackage{xcolor}
\usepackage{colortbl}

\usepackage{subcaption}
\usepackage{orcidlink}

\begin{document}

\newcommand{\qref}[1]{\textbf{RQ#1}}

\newcommand{\summaryBox}[2]{\begin{tcolorbox}[colback=black!1!white,colframe=white!20!black,center,valign=top,halign=left,before skip=2pt,after skip=5pt,center title,sharp corners,boxrule=0pt,boxsep=-2pt,left=6pt,right=6pt,width=\linewidth]\textbf{Answer to \qref{#1}:} \textit{#2}\end{tcolorbox}}

\title{Software Supply Chain Smells: Lightweight Analysis for  Secure Dependency Management}

\author{
    \IEEEauthorblockN{
        Larissa Schmid\IEEEauthorrefmark{1},
        Diogo Gaspar\IEEEauthorrefmark{1},
        Raphina Liu\IEEEauthorrefmark{1},
        Sofia Bobadilla\IEEEauthorrefmark{1},
        Benoit Baudry\IEEEauthorrefmark{2},
        Martin Monperrus\IEEEauthorrefmark{1}
    } \\
    \IEEEauthorblockA{\IEEEauthorrefmark{1}KTH Royal Institute of Technology, Stockholm, Sweden} \\
    \IEEEauthorblockA{\IEEEauthorrefmark{2}Université de Montréal, Montréal, Canada} \\
    \IEEEauthorrefmark{1}{\{lgschmid,dgaspar,raphina,sofbob,monperrus\}@kth.se} 
    \IEEEauthorrefmark{2}{benoit.baudry@umontreal.ca}
}

\pagestyle{plain}
\markboth{}{}
\IEEEpubid{}

\maketitle

\begin{abstract}
Modern software systems heavily rely on third-party dependencies, making software supply chain security a critical concern.
We introduce the concept of software supply chain smells as structural indicators that signal potential security risks.
We design and evaluate \toolname, a novel tool for detecting such smells in the supply chains of software packages.
Through interviews with practitioners, we show that our proposed smells align with real-world concerns and capture signals considered valuable. 
A quantitative study of popular packages in the Maven and NPM ecosystems reveals that while smells are prevalent in both, they differ significantly across ecosystems, with traceability and signing issues dominating in Maven and most smells being rare in NPM,  due to strong registry-level guarantees. 
Software supply chain smells support developers and organizations in making informed decisions and improving their software supply chain security posture.
\end{abstract}

\begin{IEEEkeywords}
Software Supply Chain, Open Source, Software Security
\end{IEEEkeywords}

\section{Introduction}

The number of packages provided by package managers is continuously increasing, and software projects rely on more dependencies than ever before~\cite{cox2019surviving,wittern2016look}. 
Together, these dependencies and their relationships form the \emph{software supply chain}, i.e., the network of third-party components, build tools, and distribution infrastructure involved in producing a software system~\cite{cox2019surviving,ohm_sok_2023}.
Reusing third-party libraries in software development reduces development costs by avoiding ``reinventing the wheel''~\cite{ohm_sok_2023}. Large-scale reuse also introduces new security risks~\cite{gkortzis2021reuse}, as substantial trust has to be placed in external parties. For example, installing an average NPM package implicitly means trusting 79 transitive packages and 39 maintainers~\cite{zimmermann_smallworld_2019}.

These trust relationships are often implicit and largely invisible to developers~\cite{cox2019surviving}.
As a result, many projects unknowingly depend on vulnerable or even malicious packages~\cite{drosos_bloat_2024}: Software supply chain attacks refer to attacker compromising a single dependency in order to take control of one or more downstream projects~\cite{ohm_backstabbers_2020}.

Developers need to comprehend the complex interdependencies among packages in their supply chain~\cite{ohm_sok_2023}.
This is challenging.
Packages are installed through package managers, which behave differently across programming languages \cite{gamage2025designspacelockfilespackage}. When they install dependencies as binary, these binaries do not necessarily correspond to the available source code~\cite{vu_lastpymile_2021}.
Developers have no reliable means to determine whether a dependency can be audited and trusted before including in their supply chain, or for post-mortem analysis of an incident~\cite{cox2019surviving}.

In this paper, we present \toolname, a novel approach to detect software supply chain smells and help developers securing their systems.
Extending the concept of code smells~\cite{lacerda_code_2020}, a software supply chain smell is a package that matches specific patterns indicating potential security issues, either today or in the future~\cite{liu_dirtywaters_fse}. 
For example, a dependency that does not link to its source code repository is a supply chain smell, because engineers cannot audit the source code before inclusion or cannot investigate it after a security incident.
\toolname statically analyzes three sources of information: 1) dependency files, 2) package registries, and 3) GitHub repositories, to detect software supply chain smells. \toolname  generates user-friendly reports to help stakeholders assess the quality of their software supply chains.
To evaluate the relevance and usefulness of these smells, we perform two complementary evaluations.
First, we conduct structured interviews with practitioners to gather their assessment of the smells and to understand if \toolname could integrate into existing software supply chain security practices. 
Second, we study the prevalence of these smells in practice, by conducting a quantitative analysis of the packages in the supply chain of the 50 most-depended upon packages in two ecosystems, Maven and NPM.

Our results show that \toolname captures security-relevant signals that practitioners consider important and could meaningfully strengthen existing software supply chain security processes. 
Practitioners also proposed additional indicators, especially related to project and maintainer health, that can inform future extensions of \toolname. 
Our quantitative analysis shows that a non-negligible subset of dependencies exhibits smells warranting further investigation, with clear ecosystem differences: Maven projects are more frequently affected by traceability and signing gaps -- issues practitioners rated as high or critical -- while most smells in NPM remain comparatively rare, partly due to stronger registry-level guarantees.

The concept of software supply chain smells is novel and important. State of the art software supply chain management relies on:
Software composition analysis (SCA) tools~\cite{dietrich_security_2023} that scan dependencies and match them against known vulnerability databases;
Software Bills of Materials (SBOMs)~\cite{sbom}, which provide machine-readable inventories of components and their relationships;
Security frameworks, such as in-toto~\cite{torres-arias_-toto_2019} and SLSA~\cite{slsa}, that define guidelines for securing build and release processes.
None of these tools and guidelines help developers identify dependencies that are not trustworthy, nor do they highlight which dependencies in a project's supply chain deserve closer inspection.
To the best of our knowledge, this paper provides the first systematic and empirically grounded account of software supply chain smells, combining a well-defined taxonomy, practitioner feedback, and a large-scale cross-ecosystem analysis.

To summarize, our contributions are as follows:
\begin{itemize}
    \item A novel, systematic taxonomy of software supply chain smells, informed by industry.
    \item An open-source tool, \toolname\footnote{https://github.com/chains-project/dirty-waters}, for automatically detecting those software supply chain smells across two ecosystems (Maven and NPM). \toolname can be readily integrated into modern software engineering CI/CD workflows.
    \item A qualitative evaluation of the smells and their perceived severity through interviews of 11 senior engineers.
    \item A quantitative analysis of the prevalence of software supply chain smells in the dependencies of 50 heavily depended upon projects across two ecosystems.
\end{itemize}

\section{Problem Statement}

Modern software development heavily relies on third-party components, and even small projects may depend on hundreds of external packages maintained by different parties~\cite{zimmermann_smallworld_2019}. 
A software supply chain comprises the set of libraries, tools, and infrastructure involved in developing, building, and publishing a software artifact~\cite{ohm_sok_2023}. 
This definition is recursive: the supply chain of an application is the transitive closure of the supply chains of all its dependencies.
This scale make software supply chains difficult to reason about and challenging to secure. 

Software supply chain attacks exploit the implicit trust placed in these opaque dependencies by inserting malicious code into components of the supply chain, allowing the attack to propagate to downstream consumers~\cite{ladisa_sok_2023}.
Attack vectors include the publication of malicious packages, compromise of maintainer accounts, or tampering with build and release infrastructure~\cite{williams_research_2025}. 
Due to the transitive nature of dependencies, a single compromised component can affect a large number of projects, resulting in a wide attack blast radius and delayed detection~\cite{zimmermann_smallworld_2019}.

In practice, developers have to make critical decisions that affect software supply chain security primarily in two situations: when adding a new dependency, and when updating existing dependencies. 
In both cases, developers need to assess whether the change is acceptable. 
Inspecting changes in dozens of direct dependencies is cumbersome. 
Consider transitive dependencies, which can quickly number in the hundreds and thousands:
basically developers cannot truly assess the impact of a dependency change, and are left with blind approvals of dependency changes.

\emph{The problem addressed in this paper is the lack of automated signals to evaluate the security of software dependencies.}
Adding or updating dependencies may affect reliability, but comprehensive test suites usually catch such issues. 
Security, however, remains largely invisible and existing tooling offers only limited support.
At best, dependency management tools may report known vulnerabilities in the dependencies (e.g., `npm audit`), but they provide no support in assessing the authenticity of the resulting dependency tree. 
Consequently, even with the best tools available, developers lack essential tooling regarding traceability, integrity and provenance in their software supply chains. 
Developers need effective, scalable signals to analyze dependencies for review and to make informed security decisions when accepting new dependencies or dependency version updates.

\section{Taxonomy of Software Supply Chain Smells} \label{sec:taxonomy}

The well-accepted concept of code smell refers to characteristics in the source code that indicate a potentially deeper problem, or will trigger one in the future~\cite{lacerda_code_2020}. Extending this concept, we define the concept of software supply chain smell. A software supply chain smell is a package in the dependency tree that matches specific patterns, which indicate potential security issues, current or to come in the future. 

In this section, we introduce the first taxonomy of software supply chain smells.
First, we discuss our methodology to identify the smells in \autoref{sec:taxonomy:methodology}. 
Next, we present the taxonomy of smells in \autoref{sec:taxonomy:smells}. 

\subsection{Methodology} \label{sec:taxonomy:methodology}

To identify software supply chain smells, we followed an iterative, practitioner-informed process. 
We conducted three workshops on software supply chain security with practitioners from development and security teams in Swedish companies, complemented by informal follow-up conversations. 
Based on these discussions, we derived an initial set of smells reflecting structural indicators that practitioners associate with reduced trust in dependencies. 
We then reviewed related work on dependency smells to ensure that the proposed smells were original and not already covered by existing dependency smell taxonomies.

The resulting smells were implemented in \toolname, which is available as open-source software, and piloted on real-world projects to validate feasibility and gather further feedback. 
To evaluate the relevance and severity of the smells, we subsequently conduct a practitioner study with a separate group of practitioners (cf. \autoref{sec:pract-eval}), providing empirical validation of the collected smells.

\begin{table*}
  \centering
\begin{tblr}{
  colspec = {l l  p{0.16\linewidth} p{0.16\linewidth} l p{0.16\linewidth}  }, 
  hline{1,3,12}={-}{},
  row{odd} = {gray!10},          
  row{1,2} = {font=\bfseries,white},
  cell{1}{3-Z} = {font=\bfseries},
  column{3-6} = {c},
}
\SetCell[r=2]{l} ID & \SetCell[r=2]{l} SSC Smell & \SetCell[c=4]{c} Related Software Supply Chain (SSC) Attacks \\

 &  & Distribute Malicious version & Develop Malicious Package & Inject Into Sources & Take Advantage of Vulnerabilities \\
 1 & No Source Code URL & \checkmark & \checkmark & - & - \\
 2 & Invalid Source Code URL & \checkmark & \checkmark & - & - \\
 3 & Inaccessible Release Tag & - & - & \checkmark & - \\
 4 & Deprecated & - & - & - & \checkmark \\
 5 & Fork & - & - & \checkmark & - \\
 6 & No Code Signature & \checkmark & - & - & - \\
 7 & Invalid Code Signature & \checkmark & - & - & - \\
 8 & Aliased & - & \checkmark & - & - \\
 9 & No Provenance & - & - & \checkmark & - \\
\end{tblr}
\caption{Overview of collected Software Supply Chain Smells (SSCS) and related Software Supply Chain (SSC) attacks.}
\label{tab:approach:smells}
\end{table*}

\subsection{Software Supply Chain Smells} \label{sec:taxonomy:smells}

In the following, we introduce the software supply chain smells we collected. We also outline potential security issues that each smell indicates and the attacks it enables. Table~\ref{tab:approach:smells} shows an overview of smells and related attack vectors.

\paragraph{No Source Code URL} The package's metadata does not include a URL for its associated source code repository. 
Without access to the source code, there is no insight into the code that is included and run when including this package in projects. 
This makes it impossible to audit the package for security vulnerabilities and malicious code. 
Some legitimate packages may lack public repositories, especially proprietary ones, so this smell does not always indicate malicious intent. However, if the package is meant to be open source, there is a strong expectation that its source code should be publicly available and linked from its metadata. Its absence signals poor transparency or potential concealment of malicious intent: 
This smell is a facilitator of attack vectors, "distribute malicious version of legitimate package" and "develop and advertise distinct malicious package from previous work~\cite{ladisa_sok_2023}. 

The ctx PyPI incident~\cite{ctx-pypi} is an example where linked source code enabled the detection of the distribution of a malicious version of a legitimate package. There, developers noticed that a new package release had appeared even though the linked GitHub repository had not been updated for years. This mismatch raised suspicion and enabled early detection of the compromise. The incident highlights that having a linked source code repository is essential for identifying irregularities that may signal a supply chain attack.

\paragraph{Invalid Source Code URL} The package’s metadata provides an invalid source code URL, e.g., leading to a 404 error. Similar to the previous smell, this prevents checking the source code for malicious content or vulnerabilities, enabling the same attack vectors. 
An invalid URL may be due to outdated or mistyped metadata after moving a repository. It generally indicates poor maintenance rather than malicious intent. 
However, it can also be a deliberate attempt to deceive users and simulate legitimacy by creating a false sense of transparency with a fake URL. 

\paragraph{Inaccessible Commit SHA/Release Tag} The source code repository of the package lacks the commit SHA or release tag specified in the package's metadata. 
Even though the repository provides access to the source code of the dependency, the actual version used is unknown. Without access to the release tag, it becomes impossible to trace the exact source code for a given package version and to analyze it for a malicious payload. This is of utmost importance for package registries of binary code, such as Maven, or of bundled/minified code, such as NPM.
Moreover, not knowing the exact version hinders the investigation of reasons for a security incident, as the developers cannot determine if the version of the dependency that is in the supply chain is vulnerable to a certain attack. 

The lack of the commit SHA or release tag may result from poor release management, accidental deletion or rewriting of tags.
It could also be deliberate attempts to hide unauthorized code changes. While not all of these reasons indicate malicious intent, the absence of a commit SHA or release tag weakens traceability and prevents reproducible builds of the package. 
This is connected to the attack vector "inject into sources of legitimate package": without knowing the exact version, it is impossible to know if malicious code has been injected into an otherwise authentic repository.

\paragraph{Deprecated} The package is marked as deprecated by its maintainers, either due to lack of maintenance or because of known vulnerabilities. 
This indicates that the package is no longer actively updated, making it more likely to contain unpatched security issues and exposing dependent projects to known attacks. 
Deprecation can be indicated via package metadata (on the package registry) or source repository metadata (e.g., on GitHub). 
Malicious actors have full access to this information and can exploit this not only by taking advantage of known vulnerabilities but also by taking advantage of unmaintained infrastructure, such as in the attack on the \textit{CTX} library, where the attacker was able to obtain ownership of the e-mail associated with the project maintainer because the domain had expired~\cite{ctx-pypi}.

\paragraph{Fork} 
The source code repository provided by the producer is a fork from an upstream repository.
While forks are common in open-source development, they introduce uncertainty about the authenticity of the package.
There are legitimate reasons for forks to exist, such as adding custom features, adapting code to special execution contexts, or maintaining compatibility with legacy systems. 
However, forks can also serve as an attack strategy to disguise malicious changes to the original source code and trick consumers into using compromised versions of trusted projects~\cite{cao_what_2022}:  
Attackers may introduce backdoors, add malicious installation or build steps, or replace dependencies with compromised versions while preserving much of the original code structure, relating to the "inject into sources of legitimate package" attack vector~\cite{ladisa_sok_2023}. 
When published in a package registry under a similar or misleading name, this can mislead both automated scanners and human reviewers who assume the fork is the legitimate upstream. 
A notable example is the hosting of cryptocurrency mining malware on GitHub~\cite{typosquatting-github-real}, demonstrating how forking can be abused to create deceptive repositories that appear legitimate.

\paragraph{No Code Signature} The package release is not cryptographically signed. Signatures are used to verify the authenticity of a package and can also be used to verify their integrity. Without a signature, the package's contents may have been tampered with during transmission, or the package may have been sourced from an unofficial source. 
The absence of signatures may be caused by ecosystem limitations, missing tooling support, or low awareness of signing practices, meaning the absence of a signature is not always malicious. 
But omitting a signature can also be of malicious intent to obscure a package's origin or that it has been tampered with using a Person-in-the-middle attack, relating to the "distribute malicious versions of legitimate package" attack~\cite{ladisa_sok_2023}.  

However, we note that the mere existence of a code signature is not sufficient to establish trust. 
After confirming that a package is signed, the identity of the owner of the key must be verified, and their trustworthiness must be evaluated: A valid signature only proves that the artifact was produced by a particular key, not that the key belongs to, and was used by a legitimate maintainer. This verification step is inherently non-automatable and requires human judgment or organizational trust policies.
Code signing complements provenance: while provenance attests to how and where software was built, code signatures verify who released it and that it has not been modified since.

\paragraph{Invalid Code Signature} The package release holds an invalid signature status. 
This can indicate several serious issues. It may mean that the package was tampered with after signing, that the signing key does not match the expected one, or that the signature was created using an expired or revoked key.
Malicious actors may also intentionally provide an invalid signature to give the illusion of compliance with signing requirements while avoiding verification.
An invalid signature does not have to be out of malicious intent, as it could also be due to technical errors, such as expired certificates or broken signing workflows. 
However, the presence of an invalid signature means the package's authenticity cannot be verified.
The CCleaner compromise~\cite{ccleaner} illustrates this smell: While the malicious version of the software was signed with a certificate that was valid at the time, the certificate was revoked once the breach was discovered. 
Similar to before, the mere presence of a signature is not enough to trust the package -- it is important to also verify the owner of the key.

\paragraph{Aliased} The package installs one or more of its dependencies under an alias. 
Aliasing is the practice of installing a dependency under an alternative, self-chosen name, allowing a package to install and reference different versions of the same dependency under different names. The self-chosen name is then used in the rest of the code base to refer to the dependency. 
While aliasing can be useful in some contexts, it also reduces transparency in dependency resolution and can obscure the true source or identity of a dependency, making it harder to verify which exact package is being installed and from where. 
Malicious actors may take advantage of that by re-directing alias targets to dependencies from attacker-controlled sources, compromising downstream systems without obvious signs of tampering. 
Moreover, aliasing introduces opportunities for attackers to upload malicious packages that match common alias names, exploiting users or automated systems that assume the alias corresponds to a trusted dependency. 
This smell is a realization of the attack vector "develop and advertise distinct malicious package from scratch"~\cite{ladisa_sok_2023}. 

\paragraph{No Provenance} 
A provenance file, aka build attestation, is a cryptographically signed metadata about a software artifact and its build, containing information on how and where it was built. 
A package with provenance gives substantial security guarantees about the source code origin. 
One of the ecosystems that supports provenance is NPM~\cite{npm-provenance}, allowing developers to verify where a package was built and who published a package. Sigstore~\cite{newman_sigstore_2022} is used to sign and log this information, allowing consumers to verify the authenticity and integrity of the package. 
Frameworks such as SLSA~\cite{slsa} and SSDF~\cite{ssdf} also require verifiable provenance of software artifacts to ensure that they can be trusted.

Without provenance, consumers of the package have no guarantee that the package was built from the claimed source code or by a trusted entity. 
Missing provenance enables several attack vectors. Attackers could replace the artifact or compromise build environments to inject malicious code, enabling the "inject during the build of legitimate package"~\cite{ladisa_sok_2023} attack vector.

A recent example of this is the \textit{s1ngularity} attack~\cite{s1ngularity} targeting the \textit{Nx} build system package, where attackers were able to inject code into the GitHub workflow and then publish a compromised package. Notably, the compromised package was published without provenance. 
We acknowledge that systematic provenance is an ambitious goal.
The lack of provenance today mostly stems from ecosystem limitations, missing tooling, or low awareness.
However, our research is forward looking, and, in the future, the lack of provenance will indicate efforts to conceal untrusted build environments or unauthorized code modifications.

\subsection{Discussion}

It is important to note that the absence of smells does not guarantee that a package is safe to use as the smells listed mostly target the meta-data attached to a package to assess its authenticity and origin, but do not check its contents. 
However, the absence of smells is a prerequisite for further analysis of the package security. 
Code analyses can find vulnerabilities in imported code, but if the code is not available, it becomes impossible to analyze it. 
With the large size of dependency trees and high cost of code analyses, checking for our software supply chain smells is a proxy to assess if it is worth putting more effort into analyzing a package. 
Security scanners often query vulnerability databases that contain information about vulnerabilities associated with packages and their version; however, if the commit SHA or release tag is missing in the package metadata, a user cannot be sure that the information in the database applies to the package used. 
Some organizations may have security policies that trust packages from certain organizations or maintainers; however, if no signature is provided, it is impossible to determine whether the package was indeed issued by them.

\section{\toolname{} Design \& Implementation}

In this section, we present the design of \toolname{}, a novel tool for checking software supply chain smells and informing developers in their dependency security decisions. 
The core novelty of \toolname{} is that it detects smells through the combined analysis of 1) the dependency manifest in the repository, 2) the remote package registry, and 3) the source code repository (Section~\ref{sec:approach:workflow}). 
Moreover, we design an integration of the analysis into CI/CD pipelines by designing a GitHub action to aid adoption (Section~\ref{sec:approach:action}).

\begin{figure*}
    \centering
    \includegraphics[width=1\linewidth]{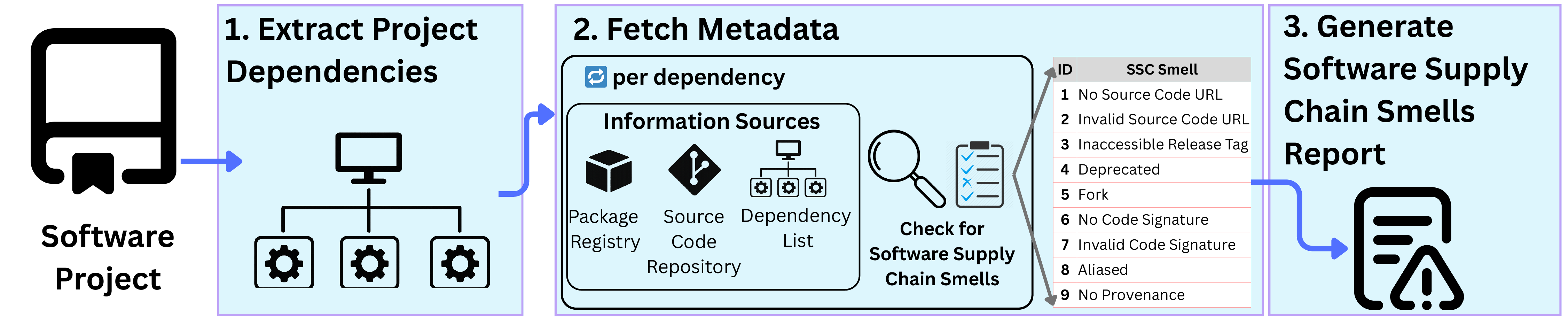}
    \caption{Workflow of \toolname{}. Based on the dependency tree of a project, it gathers analytics from the dependencies and the package registries, to generate a software supply chain smells report. 
    }
    \label{fig:overview}
\end{figure*}

\subsection{Workflow} \label{sec:approach:workflow}

Figure~\ref{fig:overview} shows an overview of the workflow of \toolname{}. \toolname{} receives as input the project name and version to be analyzed, and the package manager it uses. 
The workflow consists of three stages. 
First, the tool extracts the complete dependency list, including both direct and transitive dependencies, from the project's dependency manifest, whose format depends on the package manager used by the project.
Next, it fetches and accumulates metadata for each extracted dependency from several sources. 
For this, the tool starts with fetching metadata from the package registry. 
If that contains a valid GitHub repository URL, smells pertaining to the repository are checked (Inaccessible Release Tag, Fork). Information about aliasing, i.e., if and which dependencies are installed under alternative names to allow referencing different versions of the same package (cf.~\ref{sec:taxonomy:smells}), is extracted from the dependency list. 
The dependency list explicitly defines aliases, mapping an alias name to a specific package source or version.
Finally, the collected data is stored in a file that will be later used to generate the summary of software supply chain smells.

\subsubsection{Extract Project Dependencies}

To extract both direct and transitive dependencies, \toolname{} analyzes the given project's dependency manifest. \toolname{} is independent of the specific ecosystem and relies on the respective package manager to provide the complete, resolved list of dependencies. 
For NPM, the lockfile is used, as it records the exact versions and sources of all dependencies, including transitive ones. Unlike the dependency manifest, which specifies version ranges or constraints, the lockfile captures the resolved, concrete dependency tree. 
For Maven, we utilize the \texttt{maven-dependency-plugin} for parsing the \texttt{pom.xml} to retrieve the complete dependency tree, incl. the transitive ones. 

\subsubsection{Fetch Metadata}

\toolname{} retrieves various metadata according to the different smells introduced in Section~\ref{sec:taxonomy:smells}. 
As shown in Figure~\ref{fig:overview} and Table~\ref{tab:tool:smells-support}, \toolname{} utilizes different sources to analyze their prevalence: 
The package registry is queried for the source code URL, deprecation status, code signature, and provenance file. 
To check the accessibility of the source code URL, \toolname{} sends a request and verifies the HTTP response. 
If the link is accessible, \toolname{} then assesses whether the repository is a fork from data in the source code repository. Moreover, the accessibility of the release SHA is assessed; if none is present or it is found to be invalid, the accessibility of the release tag is assessed based on the common tagging patterns identified by Keshani et al.~\cite{keshani2024aroma}. 
To check if a package is aliased, \toolname{} utilizes the extracted dependency list from the project that is checked, as the dependency list contains information about aliases used directly and in transitive dependencies. 
The collected data is stored in a JSON file for report generation, which we call the "Dirty Pond".

Due to constraints in Maven's ecosystem and its limited package metadata, not all smell checks are supported for Maven projects: namely, for deprecation and provenance.
Moreover, aliasing is not supported by Maven. 
Table~\ref{tab:tool:smells-support} shows an overview of the supported smell checks by package manager. 

\subsubsection{Generate Software Supply Chain Smells Report}

Utilizing the Dirty Pond, \toolname{} generates a comprehensive, human-readable Markdown summary
of the analysis. 
The goals of the report are to raise awareness of supply chain smells, support informed decision-making and priorization of issues, and improve transparency about the trustworthiness of dependencies. Therefore, the report has to prioritize clarity and readability to make results understandable to developers. This includes ensuring actionability by outlining next steps for mitigation, maintaining traceability by linking each smell to the affected packages, and providing contextual explanations that describe why each issue matters and what risks it introduces.

\begin{figure}
    \centering
    \includegraphics[width=0.8\linewidth]{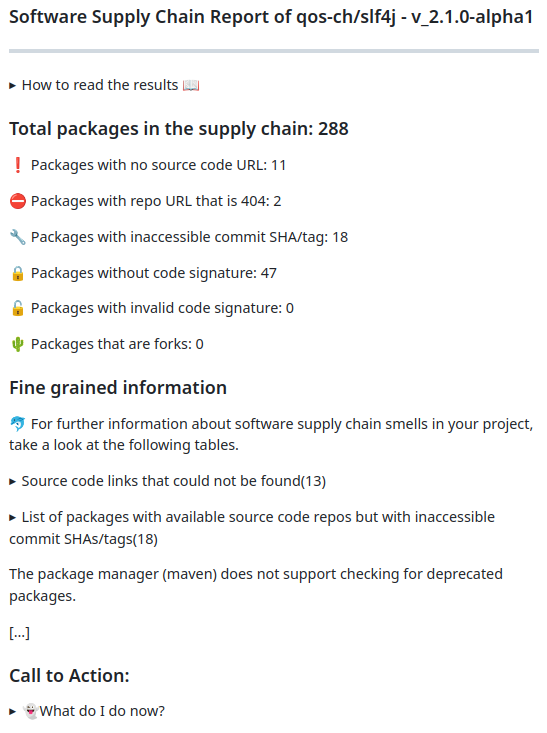}
    \caption{Excerpt of a Software Supply Chains smells report generated by \toolname{}.}
    \label{fig:example-report}
\end{figure}

Figure~\ref{fig:example-report} shows an excerpt of a report generated by \toolname{} for the \texttt{qos-ch/slf4j} package: 
The report begins with instructions on how to interpret the results, followed by key points related to the software supply chain smells uncovered by \toolname{} and why they matter together with a list of packages that expose each smell.
Importantly, the report includes a "Call to Action" section to provide guidelines for developers to fix the smells: 

\paragraph{No or Invalid Source Code URL, Inaccessible Commit SHA/Release Tag} Consumers should submit a Pull Request to the dependency's maintainer, requesting correct repository metadata and proper tagging.
\paragraph{Deprecated} Confirm the maintainer’s deprecation intention and
double-check for alternative versions that are not deprecated.
\paragraph{Fork} Inspect the package and its GitHub repository to verify
that the fork is not malicious.
\paragraph{No Code Signature} Open an issue in the dependency's repository to request the inclusion of code signature in the CI/CD pipeline. 
\paragraph{Invalid Code Signature} Verify the code signature and contact the maintainer to fix the issue.
\paragraph{Aliased Package} Check the aliased package and its repository to verify that the alias is not malicious. 
\paragraph{No Provenance} Open an issue in the dependency’s repository to request the inclusion of provenance and build attestation in the CI/CD pipeline. Note that, while this is still cutting-edge today, we believe that this is the future of software supply chain security.

\subsection{Integration in Continuous Integration} \label{sec:approach:action}

It is crucial to integrate static analysis tools into modern software engineering workflows to foster their adoption~\cite{smith_why_2020}. 
Therefore, we design \toolnameaction{}, a GitHub action that can be integrated in GitHub workflows CI/CD pipelines to provide automatic feedback to project developers and maintainers on the security of their software supply chain. 
This idea is that when the CI is triggered, \toolname{} is automatically executed after checking out its repository at a certain version and installing its dependencies.

\textbf{Failing the build}. \toolnameaction{} can fail the CI if high-severity smells are found. 
This enforces immediate attention to critical supply chain issues and ensures that developers address them before merging or deployment. 
By failing the build and making supply chain security a mandatory check in the build process, \toolnameaction{} elevates software supply chain security from a secondary consideration to a first-level concern for developers. 
While this enforces stricter security standards, the configuration of \toolnameaction{} allows balancing security enforcement with development flexibility by triggering failures only for high-severity or threshold-exceeding smells, while less critical issues are shown as informative reports.

\textbf{Automated reporting}.  \toolnameaction{} publishes the software supply chains report generated as comment to the pull request or commit, depending on the configuration. This raises awareness by clearly showing which actual smells exist in the supply chain, helping developers understand risks early and enabling them to assess whether they want to proceed with the change or review it and reduce the number of smells before merging. 
Attaching the report directly to the PR or commit also provides traceability and documentation, allowing teams to track issues over time and review the history of supply chain problems and their resolution.

\definecolor{nocolor}{HTML}{949698} 
\newcommand{\n}{{\textcolor{nocolor}{$\times$}}}

\begin{table}
  \centering
\rowcolors{3}{gray!10}{white}
  \begin{tabular}{l l c c c c c}
    \toprule
    \textbf{ID} & \textbf{SSC Smell} & \textbf{P} & \textbf{S} & \textbf{D} & \textbf{NPM} & \textbf{Maven} \\
    \midrule
    1 & No Source Code URL       & \checkmark &            &            & \checkmark & \checkmark \\
    2 & Invalid Source Code URL  & \checkmark &            &            & \checkmark & \checkmark \\
    3 & Inaccessible Release Tag &            & \checkmark &            & \checkmark & \checkmark \\
    4 & Deprecated               & \checkmark &            &            & \checkmark & \n \\
    5 & Fork                     &            & \checkmark &            & \checkmark & \checkmark \\
    6 & No Code Signature        & \checkmark &            &            & \checkmark & \checkmark \\
    7 & Invalid Code Signature   & \checkmark &            &            & \checkmark & \checkmark \\
    8 & Aliased                  &            &            & \checkmark & \checkmark & \n \\ 
    9 & No Provenance            & \checkmark &            &            & \checkmark & \n \\
    \bottomrule
  \end{tabular}
  \caption{Overview of supported Software Supply Chain Smells (SSCS) by package manager and information source: \textbf{P} = package registry, \textbf{S} = source code repository, \textbf{D} = dependency list. 
  Supported checks indicated by \checkmark, checks not supported by the package manager by \n.}
  \label{tab:tool:smells-support}
\end{table}

\section{Practitioner Study: Evaluation of \toolname Smells} \label{sec:pract-eval}

To assess the relevance of \toolname(cf. Section~\ref{sec:taxonomy:smells}), we conduct a campaign of structured interviews with qualified practitioners to answer the following research questions: 
\begin{enumerate}[label=\textit{RQ\arabic*:},leftmargin=27pt]
    \item How severe do practitioners consider \toolname smells? 
    \item What additional smells do practitioners consider important? 
\end{enumerate}

First, we describe the methodology of our user study in Section~\ref{sec:pract-eval:meth}. 
We then report the results and discuss them in two parts: first, practitioners' severity ratings of our proposed smells (Section~\ref{sec:pract-eval:rating}), and second, the additional smells they suggested (Section~\ref{sec:pract-eval:additional-smells}).

\subsection{Methodology} \label{sec:pract-eval:meth}

To evaluate how serious practitioners consider the proposed smells, we conducted interviews with industry professionals who have experience regarding software supply chain security. This included developers, team leads, and others involved in dependency-related engineering decisions. We contacted potential participants by e-mail, selecting them from a mailing list they had previously subscribed to, which we use to share news about our software supply chain research. The e-mail included a short introduction to our research and an outline of the interview.

After they agreed to take part and signed a consent form, we held semi-structured interviews on Zoom, each lasting about 30 minutes. During the interviews, we first asked participants to introduce themselves. We then introduced the \toolname smells introduced in Section~\ref{sec:taxonomy:smells} to them one by one and asked them to rate the severity of each of the smells regarding their impact on security, regarding maliciousness and vulnerability. We also asked about a short reasoning why they felt the ranking was appropriate. Next, they were asked to mention any smells they felt were missing. Finally, they were given the opportunity to describe their current practices for handling software supply chain security. 
We recorded each interview and transcribed it to text using Whisper\footnote{\url{https://github.com/openai/whisper}}.

In total, we interviewed eleven practitioners (cf. Table~\ref{tab:practitioners}). Eight of them had more than twenty years of professional experience, two had eighteen years, and one had around ten years. Nine participants worked in security, while one focused on quality assurance and one was a development manager. Their technical backgrounds were diverse: four primarily worked with Java, two with Go and Python, respectively, and one each with C/C++, .NET, and Rust. 
Companies they worked for included, but are not limited to Google, Oracle, Sonatype, and Keyfactor. The opinions shared by the interviewees, however, do not represent an official statement by these companies. 

\begin{table}
\centering
\begin{adjustbox}{width=.95\linewidth}
\rowcolors{3}{gray!10}{white}
\begin{tabular}{lrll}
\toprule
\textbf{ID} & \textbf{YoE} & \textbf{Sector} & \textbf{Tech Stack} \\
\midrule
P1 & $>20$ & IT Consulting & \textbf{C/C++}, Linux \\ 
P2 & $18$ & Information Technology & \textbf{Java}\\
P3 & $18$ & Game Development & \textbf{.NET} \\
P4 & $>20$ & Consulting -- Public Sector & \textbf{Java}, JavaScript \\ 
P5 & $>20$ & Information Technology & \textbf{Java} \\
P6 & $>20$ & Information Technology & \textbf{Go}\\
P7 & $>20$ & Information Technology & \textbf{Go}, Linux (Debian) \\
P8 & $>20$ & Information Technology & \textbf{Java} \\
P9 & $>20$ & Information Technology & \textbf{Python}, Java \\
P10 & $10$ & Information Technology & \textbf{Rust}, Python, Node.js, Nix \\
P11 & $20$ & Information Technology & \textbf{Python} \\
\bottomrule
\end{tabular}
\end{adjustbox}
\caption{Years of experience, sector, and tech stack of the interviewees. Primary tech stack used in \textbf{bold}.}
\label{tab:practitioners}
\end{table}

\subsection{RQ1: Rating of Smells}\label{sec:pract-eval:rating}

In this section, we present and discuss how practitioners rate the \toolname smells introduced in Section~\ref{sec:taxonomy:smells}. 

\subsubsection{Results}

Table~\ref{tab:smell_ratings} shows how the interviewed practitioners rated the severity of our software supply chain smells. 
Six out of nine smells received at least one rating as critical. 
\textbf{Invalid code signature} showed strong consensus as a critical smell with eight participants rating it as critical. Participants mentioned that an invalid signature is a sign of broken integrity and \textit{"that something is not as it should be" (P10)} to them. One participant (P7), however, classified this smell as low, noting that the signature could become invalid due to an expired key, which would not bother them. 
\textbf{Invalid source code URL} was rated as critical by four participants, and as high by another three. Participants reasoned that this signals that the code cannot be trusted and can be a blocker for using it: \textit{"You cannot trust the source code" (P2)}, \textit{"[I] would like to know what the source code is and I cannot get the answer to that question" (P7)}, \textit{"that would be a blocker" (P4)}. 
Similarly, \textbf{without source code URL} received three ratings as critical and another five as high, with participants noting that there should be no reason to not give the URL in open source development, and that they would therefore not trust the package; for example, P2 noted \textit{"I would not trust that (...) and I would not use such a package"}, and P2 stated that \textit{"we should always require that the source code is available"}. 
Participants also perceived the \textbf{Inaccessible commit SHA/tag} smell as serious, with two ratings as critical and five as high, mentioning that knowing the exact version of a package is a prerequisite for reproducibility of builds (P5) and that not having it could indicate that \textit{"someone is tampering with the package"} (P4). P6 stated that this smell is \textit{"not as bad as no source code at all, but still bad"}. 

\textbf{No code signature} received more mixed ratings: 2 participants rated it as critical, and another four as high, reasoning that they cannot be sure by whom the package is distributed and if the code is not tampered with. 
Another four participants, however, rated the smell as low or medium; three because they would not expect to see code signatures in their ecosystem (P3/.NET, P8/Java, and P11/Python), one stressing that the value of a code signature also \textit{"depends on the ability to check by whom it is signed" (P7)}. 
The \textbf{deprecated} smell was mostly rated between medium and high. Participants rating the smell as critical noted that \textit{"if you have it [the dependency] and it is deprecated, then you should invest work on getting rid of this dependency" (P8)}, and that they are not allowed to use it at all (P2). Participants rating the smell as medium mentioned a trade-off between the additional work of switching to another package and security risks, which they would look at on a case-by-case basis.

\textbf{Forks} were rated as medium by seven out of eleven participants because they are common and legitimate in open-source development. Yet, eight practitioners still expressed that a forked dependency warrants closer inspection to understand its purpose and legitimacy; for example, P2 noted that \textit{"we would clearly look into it and try to figure out why they did the fork"} and that \textit{"we need to to dig into these things fairly in detail before using some third-party dependency"}.
For \textbf{Aliases}, two participants reasoned that they should be avoided for reasons of transparency, but do not pose a security risk itself. Another participant mentioned relying on security scanners for a problem like that, while another did not judge aliasing as problematic as long as the alias name of dependencies cannot be used in their own code base: \textit{"unless you can use the very same alias in your own code base without defining it yourself (...) I would not look at the aliases" (P10)}. 

\textbf{No Provenance}, i.e., not providing cryptographically signed metadata about a software artifact and its build (cf. Section~\ref{sec:taxonomy:smells}), was rated low or medium by nine out of eleven practitioners. This is not because they thought of it as not important, but because it is not yet widely adopted in practice. Several participants noted that they would like to see more packages providing provenance, stating for example, that \textit{"this would be great"} (P8), and would therefore like to ideally rate this smell higher.

Some participants gave multiple ratings for the same smell feeling that they are context-dependent; for clarity, we classified them conservatively as the respective lower category. 
One participant classified the inaccessible commit SHA/release tag as "between medium and high", another participant rated the absence of a tag as high, but the absence of a commit SHA as critical.
Similarly, one participant rated the deprecated smell as medium for "packages with small attack surfaces", and as high for packages with large attack surfaces. Another made the rating of the fork smell dependent on whether the original project was still active (low rating) or not (critical rating). 
One participant did not rate the provenance smell, as it does not apply to their context, where they are building all software from source code themselves.
Another participant did not rate both the no code signature and invalid code signature smell, stating that they are not well-phrased, as the smell should be about checking the owner of the signing key, not if there is a signature. 
Five participants did not feel comfortable providing a rating for the Alias code smells, as the ecosystems they work with do not support aliasing.

\subsubsection{Discussion}\label{sec:pract-eval:discussion:rating}

The interview results demonstrate that our proposed software supply chain smells are meaningful and align well with practitioners' real-world concerns. 
Most smells were rated as critical by at least one participant. In particular, the smells related to source authenticity and signature validity received consistently high or critical ratings. 

Severity assessments often depended on the participant's background and ecosystem; generally, participants rated smells as low not only if they thought it was not a meaningful indicator for a security risk, but also if it did not apply to their particular context. 
For example, one participant rated the Inaccessible Release Tag smell as medium because they build the source code themselves. Similarly, two participants rated the No Code Signature smell as low and medium, respectively, arguing that signatures are not a "must have" due to not being widely adopted yet. 

Some practitioners emphasized that smells cannot be judged in isolation. Multiple weak indicators could compound, making the combined risk greater than the sum of individual smells. Others also stressed that changes across versions -- such as disappearing provenance, altered signing behavior, or unexpected repository moves -- would immediately raise suspicion, sometimes more than a static absence of data. 
Several participants further highlighted that the absence of a smell does not guarantee that a package is secure to use, reinforcing our view that the smells are best viewed as indicators that reveal fundamental reasons not to trust a package.

More broadly, participants noted that they are cautious about introducing new dependencies. Integrating smell detection into existing governance processes could provide actionable transparency to enable informed decisions about new dependencies without adding heavy manual effort. 
Moreover, several participants mentioned gaps in continuous monitoring of dependencies, noting that dependencies are evaluated when first introduced but not consistently re-assessed when updated. \toolnameaction{} could help fill this gap by offering lightweight, repeatable checks at every update.

\summaryBox{1}{The results from the interviews with expert developers show that our smells capture signals practitioners consider valuable. The collected evidence shows that the proposed smells are grounded in practical needs and  inform supply chain security decisions.}

\begin{table}
\centering
\begin{adjustbox}{width=.95\linewidth}
\rowcolors{3}{gray!10}{white}
\begin{tabular}{llrrrrr}
\toprule
\textbf{ID} & \textbf{Smell} & \textbf{L} & \textbf{M} & \textbf{H} & \textbf{C} & \textbf{N} \\
\midrule
1 & No Source Code URL & 1 & 2 & \textbf{5} & 3 & 0 \\
2 & Invalid Source Code URL & 1 & 3 & 3 & \textbf{4} & 0 \\
3 & Inaccessible Release Tag & 1 & 3 & \textbf{5} & 2 & 0 \\
4 & Deprecated & 1 & \textbf{4} & \textbf{4} & 2 & 0 \\
5 & Fork & 3 & \textbf{7} & 1 & 0 & 0 \\
6 & No Code Signature & 2 & 2 & \textbf{4} & 2 & 1 \\
7 & Invalid Code Signature & 1 & 0 & 1 & \textbf{8} & 1 \\
8 & Aliased & 3 & 2 & 1 & 0 & \textbf{5} \\
9 & No Provenance & \textbf{7} & 2 & 1 & 0 & 1 \\
\bottomrule
\end{tabular}
\end{adjustbox}
\caption{Practitioner ratings of software supply chain smells; \textbf{L} = low severity, \textbf{M} = medium, \textbf{H} = high, \textbf{C} = critical, \textbf{N} = no rating. Most common rating in \textbf{bold}. The high severity validates \toolname's relevance.}
\label{tab:smell_ratings}
\end{table}

\subsection{RQ2: Additional Smells} \label{sec:pract-eval:additional-smells}

In this section, we present additional smells proposed by practitioners and if and how they could become part of the smells checked in \toolname.  

\subsubsection{Results}

All participants suggested one or more additional smells they considered relevant for assessing the security of a package before being added as a dependency. Table~\ref{tab:additional_smells} shows an overview of proposed smells. 
Many interviewees highlighted project and maintainer health, pointing to indicators such as contributor and maintainer activity and identity, i.e., who maintainers are (mentioned by six interviewees), long periods without updates (three mentions), the age of a package (one mention), irregular release patterns (one mention), and missing security testing (one mention). 
Others focused on the dependency footprint and usage, noting that an overly complex dependency tree (four mentions), limited real use of a dependency (one mention), or unclear adoption levels by other users (two mentions) can signal risk to them. 
Several suggestions concerned authenticity and trust, including checks for suspicious name similarity (two mentions), verification of signature owners on top of checking for a valid signature, unexpected changes in project location or build setup, or cases where a repository appears to be a copy rather than a proper fork (one mention each). 
Interviewees also mentioned build and release integrity issues, such as missing reproducible builds (two mentions) or binaries bundled within source releases (one mention). 
Finally, a few pointed to behavior and permission-related smells, for example, packages requesting many access rights (one mention) or exhibiting unexpected network behavior during install or runtime (one mention).

\subsubsection{Discussion} \label{sec:pract-eval:discussion:additional-smells}

The additional smells proposed by practitioners cover a wide range of potential indicators for issues.
Several of these suggestions could be integrated into future extensions of \toolname (cf. Table~\ref{tab:additional_smells}), as they align with our design goal of supporting smells that can be extracted reliably from metadata. For example, long periods without updates or the age of a package could be derived from release timestamps, while adoption levels could be approximated using download statistics from package registries. Similarly, suspicious name similarity -- often associated with typosquatting attacks -- could be detected using package naming metadata and existing tooling for detecting potential typosquatting packages~\cite{typosquatting-tool}. Some build-related suggestions, such as checking for reproducible builds or binaries bundled within source releases, could also be incorporated, although they may introduce trade-offs in terms of analysis time and are more ecosystem-specific (e.g., particularly relevant for Maven/Java).

Several other suggestions, however, fall outside the scope of our current analysis. Indicators related to project and maintainer activity and identity, irregular release patterns, or the "reasonableness" of a dependency tree are typically gradual and highly contextual, making them difficult to express as generalized smells that are either present or absent. Likewise, signals such as missing security testing, verification of signature owners, unexpected changes in project location or build setup, or distinguishing copied repositories from legitimate forks are challenging to extract automatically and reliably at scale. Finally, smells based on runtime behavior or requested access rights require executing the software and are therefore incompatible with our static, metadata-based approach.

The interview results also show a broader set of practical needs. 
Participants expressed interest in understanding the amount of additional transitive dependencies included when using a certain package, sometimes in combination with understanding how much of a package is used in their code, which does not translate into a binary smell. 

To our knowledge, this is the first detailed account in academic literature of how practitioners themselves evaluate supply chain smells and what additional signals they consider relevant. Our findings provide novel insight into real-world needs and expectations that existing research has not systematically captured.

\summaryBox{2}{Practitioners also care about other software supply chain smells related to project  health and maintainers identity. The proposed smells call for future research and development based on our grounded findings.}

\begin{table}
\centering
\begin{adjustbox}{width=1\linewidth}
\begin{tabular}{llcl}
\toprule
\textbf{Category} & \textbf{Proposed smell} & \textbf{Mentions} & \textbf{In scope?} \\
\midrule
\multirow{5}{*}{Project \&  maintainer health}
  & Maintainer identity/activity & 6 & \n \\
  & No recent updates & 3 & \checkmark \\
  & Old package & 1 & \checkmark \\
  & Irregular releases & 1 & \n \\
  & No security testing & 1 & \n \\
\midrule
\multirow{3}{*}{Dependency footprint}
  & Large dependency tree & 4 & \n \\
  & Unclear adoption & 2 & \checkmark \\
  & Low real-world usage & 1 & \checkmark \\
\midrule
\multirow{4}{*}{Authenticity \& trust}
  & Suspicious name similarity & 2 & \checkmark \\
  & Project location/build changes & 1 & \n \\
  & Copied repository & 1 & \n \\
  & Unverified signature owner & 1 & \n \\
\midrule
\multirow{2}{*}{Build \& release integrity}
  & No reproducible builds & 2 & Future \\
  & Binaries in source release & 1 & Future \\
\midrule
\multirow{2}{*}{Behavior \& permissions}
  & Excessive permissions & 1 & \n \\
  & Unexpected network behavior & 1 & \n \\
\bottomrule
\end{tabular}
\end{adjustbox}
\caption{Additional software supply chain smells proposed by practitioners, grouped by category. }
\label{tab:additional_smells}
\end{table}

\section{Quantitative Study Across Ecosystems} \label{sec:eval-ecosystems}

In this section, we present a quantitative analysis of software supply chain smells across two package ecosystems. Our goal is to understand how common the smells are in practice and how their prevalence differs between ecosystems. We examine the following research question: 

\begin{enumerate}[label=\protect\textbf{RQ\arabic*}, leftmargin=*]
    \setcounter{enumi}{2}
    \item\label{RQ3} What smells do packages in the supply chain of popular packages in different ecosystems exhibit? 
\end{enumerate}

\noindent
First, we describe our methodology for collecting and analyzing packages from each ecosystem in Section \ref{sec:eval-ecosystems:methodology}. We then report our findings in Section \ref{sec:eval-ecosystems:RQ4}, followed by a discussion of threats to validity in Section \ref{sec:eval-ecosystems:ttv}. 

\subsection{Methodology} \label{sec:eval-ecosystems:methodology}

We analyze the 50 most depended-on packages in the NPM and Maven ecosystems.
We consider a project if it is hosted on GitHub and:
1) for NPM, has a \textit{package-lock.json} file,
2) for Maven, has a \textit{pom.xml} file. We use the metric of most depended-on packages as a proxy for ecosystem impact as retrieved by npm-rank\footnote{https://github.com/tristan-f-r/npm-rank} for NPM, and the libraries.io\footnote{https://libraries.io/} service for Maven. For each package, we retrieve its GitHub repository path and its most recent version as of November 25, 2025, and the matching release commit or release tag. We use a package's most recent version to ensure that we do not report issues that have already been fixed at the time of analysis.

We run \toolname until we obtain fifty successfully completed analyses. 
For Maven, nine analyses fail, seven due to an error in the \textit{maven-dependency-plugin} when resolving dependencies, and two due to an error in the \textit{resolve-plugin}. 
For NPM, nine analyses fail due to non-supported formatting of lockfiles. 
We provide the complete list of packages analyzed together with the analysis results as part of our supplementary material~\footnote{\url{https://github.com/chains-project/dirty-waters-utils/tree/master/scripts/smell-presence-analysis/2025-11}.}.

To answer RQ3, we consider all individual packages that appear in the dependency trees of the 50 analyzed projects per ecosystem. For each package, we analyze whether it directly exhibits any of the defined software supply chain smells, independent of the project that depends on it. This allows us to study how common each smell is among packages in the ecosystem and to compare the distribution of smells between Maven and NPM. 

\subsection{RQ3: Direct smells in packages} \label{sec:eval-ecosystems:RQ4}

\begin{figure}
    \centering
    \includegraphics[width=0.99\linewidth]{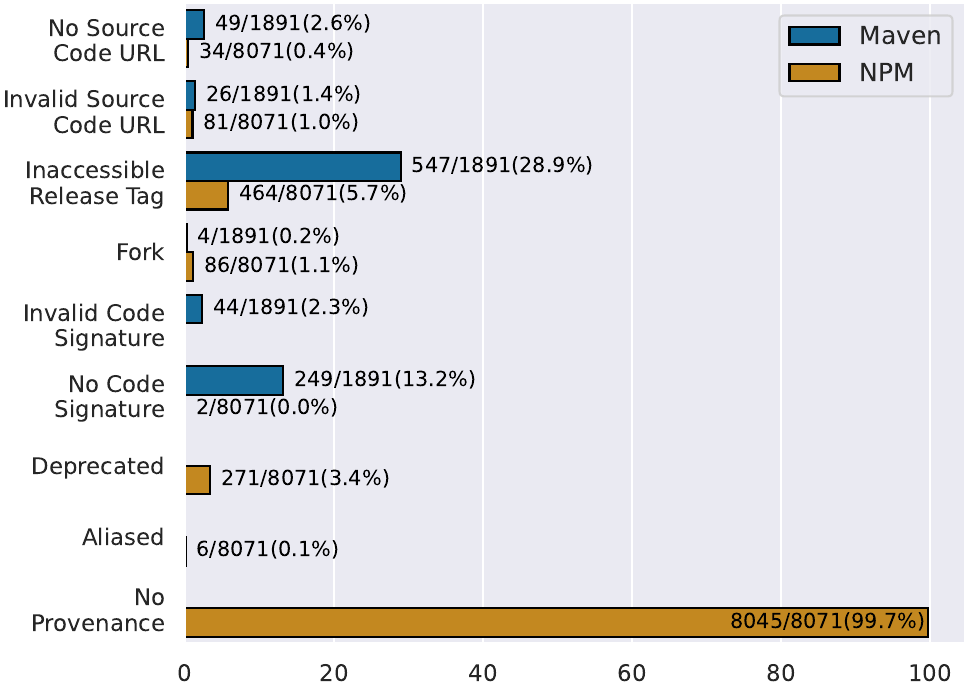}
    \caption{Distribution of smells exhibited by packages directly. Note that the Deprecated, Aliased, and No Provenance smells are not supported for Maven.}
    \label{fig:all-packages-smell-distribution}
\end{figure}

Figure~\ref{fig:all-packages-smell-distribution} summarizes how many packages exhibit each smell directly, considering all packages present in the dependency trees of the top 50 projects per ecosystem. This amounts to 1891 unique packages for Maven and 8071 for NPM. 
Knowledge about smells is particularly relevant when adding a new dependency or updating a dependency version, as they indicate whether a package provides sufficient traceability, integrity, and provenance, supporting informed security decisions.

Let us first discuss the smells for Maven (blue bars). 
The most frequent direct smell is \textbf{inaccessible commit SHA}, affecting 547 packages (62.8\%). This is significant, as practitioners frequently rated this smell as high or critical due to the resulting loss of traceability and auditability. The important presence of this smells is evidence that linking released artifacts back to specific source code versions remains a common challenge in the ecosystem. When adding a dependency with this smell, developers cannot reliably trace the released artifact back to a specific source code version, limiting their ability to review changes or respond to incidents. 
The second most common smell is \textbf{no code signature} (249 packages, 28.6\%). Given that Maven has long supported artifact signing~\cite{maven-signing}, this represents a substantial fraction of dependencies lacking integrity and authenticity guarantees. For a developer evaluating a new dependency, this means there is no cryptographic assurance that the artifact originates from the expected publisher. 
Other smells are considerably less common: 
\textbf{No source code URL} (49 packages, 5.6\%), \textbf{invalid code signature} (44 packages, 5.1\%), 
\textbf{invalid source code URL} (26 packages, 3\%), and \textbf{forked} projects (4 packages, 0.5\%).

For NPM (orange bars), most direct-package smells appear far less frequently in relative terms, though absolute counts remain at around the same numbers compared to Maven due to the larger dataset. 
Nearly all packages in the dependency tree exhibit the \textbf{no provenance} smell (8045 packages, 99.7\%). This aligns with our expectation that the smell is forward-looking, as provenance is not yet widely adopted. 
The second- and third-most common direct smells are \textbf{inaccessible commit SHA} (464 packages, 5.8\%) and \textbf{deprecated} (271 packages, 3.4\%). When adding a dependency, these smells signal potential difficulties in reviewing source code or relying on long-term maintenance, respectively. 
\textbf{Invalid source code URL} (81 packages, 1.0\%) and \textbf{forked} Projects (86 packages, 1.1\%) also occur but remain rare overall. 
Smells related to signing are nearly absent: \textbf{No code signature} (2 packages, 0.0\%) and \textbf{invalid code signature} (0 packages). 
As packages in NPM are signed by the registry\footnote{https://docs.npmjs.com/about-registry-signatures}, it is normal to find no packages without or with an invalid code signature, highlighting how ecosystem design shapes the supply chain risk profile and developers adding new dependencies in NPM benefit from strong registry-level integrity guarantees by default.

\summaryBox{3}{For Maven, the most common smells are related to traceability and missing code signatures, smells that were mostly rated as high or critical by practitioners, suggesting that Maven would benefit from stronger registry-level guarantees.
For NPM, all smells remain relatively rare, demonstrating strong ecosystem awareness to dependency problems.}

\subsection{Threats to Validity} \label{sec:eval-ecosystems:ttv}

We discuss how we address threats to validity as outlined by Wohlin et al.~\cite{Wohlin2012} and Runeson and Höst~\cite{runeson2008}.

\paragraph{Internal Validity} A first threat concerns the use of the 50 most depended-upon projects per ecosystem as a proxy for popularity. While this excludes less common packages, we are confident that our analysis still provides a reasonable view of typical supply chain practices because these projects anchor large parts of the ecosystem. 
Tool failures (nine Maven, nine NPM) present another threat, but since our goal is to capture ecosystem-level tendencies rather than evaluate specific projects, we do not expect these missing cases to affect the overall patterns.
Dependency resolution and repository mapping may introduce inaccuracies, especially in Maven multi-module projects where we analyze the full repository. These effects apply broadly and are unlikely to skew comparative results. 
Reliance on metadata fields also creates a risk of misclassification, as metadata may be incomplete or inconsistent. However, since such inconsistencies are themselves supply chain concerns, detecting them aligns with the purpose of our analysis. 

\paragraph{External Validity} Our results may not generalize to the full ecosystems, as we sampled only the most depended-upon packages, which are typically better maintained. However, because these projects influence many downstream consumers, they still offer valuable insight into common dependency patterns. A further limitation is the requirement for a lockfile or standard Maven setup, and the restriction to GitHub-hosted repositories, which excludes some projects. 
Given the prevalence of GitHub and standard build practices in widely used packages, we expect this to have a limited effect on generalizability.
\section{Related Work}

\subsection{Software Supply Chain Attacks}

Prior work has proposed taxonomies to classify software supply chain attacks and their attack vectors. 
Ohm et al.~\cite{ohm_backstabbers_2020} derive a taxonomy from 174 malicious open-source packages. 
Ladisa et al.~\cite{ladisa_sok_2023} present an ecosystem-agnostic attack taxonomy covering all stages of open-source supply chains. 
Gokkaya et al.~\cite{gokkaya2026software} extend this line of work by linking attack vectors to mitigations and present a framework for software supply chain risk assessment.

Beyond taxonomies, several approaches aim to detect active software supply chain attacks. 
Tan et al.~\cite{tan2026operationalruntimebehaviormining} mine anomalous runtime behavior to identify compromised dependencies. 
Zhen et al.~\cite{zheng2024towards} present OSCAR, which detects dynamic code poisoning by fully executing packages in sandboxed environments and monitoring behavior through fuzzing and API hooks. 
Sejfia et al.~\cite{sejfia2022practical} apply machine learning to classify malicious packages based on observed characteristics, and Reyes et al.~\cite{reyes2025mavenhijacksoftwaresupplychain} study a specific attack technique in Maven exploiting class resolution order at runtime.

In contrast, our work does not detect concrete attacks. We focus on structural indicators -- software supply chain smells -- that undermine trust and transparency in dependencies and increase susceptibility to the attack vectors identified in prior work.

\subsection{Vulnerability-Centric Supply Chain Analysis}

Much prior work focuses on identifying known vulnerabilities in software supply chains. 
Software Composition Analysis (SCA) tools generate inventories of third-party components and check them against vulnerability databases~\cite{imtiaz2021comparative}, while static analysis techniques aim to detect vulnerable code patterns directly in source code~\cite{hastings_continuous_2022}. 
Comparative studies show that SCA tools vary significantly in their effectiveness, largely due to differences in the accuracy of queried vulnerability databases~\cite{imtiaz2021comparative}.

Several studies highlight the limitations of vulnerability-centric approaches. 
Imtiaz et al.~\cite{imtiaz2023open} report a median delay of 17 days between the release of a security fix and the publication of a corresponding security advisory. 
Mir et al.~\cite{mir2023on} apply reachability analysis to Maven dependencies and find that less than 1\% of packages have reachable call paths to vulnerable code, indicating that impossible reachability is a major cause of false positives. Other work analyzes vulnerable transitive dependencies at scale using CVE data~\cite{mahon2025pypitfall,marquez_dataset_2025,german_marquez_vulnerability_2024}.

While these approaches center on known vulnerabilities and disclosed security issues, our work targets properties of packages and their metadata that affect trust and auditability independently of known vulnerabilities, enabling risk assessment even in the absence of CVEs or security advisories.

\subsection{Metadata- and Repository-Based Risk Indicators}

Prior work has explored metadata- and repository-based indicators to assess risks in software supply chains, including repository accessibility~\cite{tsakpinis_analyzing_2024}, malware injection in artifacts~\cite{vu_towards_2020}, maintainer activity~\cite{zahan_what_2022}, repository popularity~\cite{borges_understanding_2016}, and malware in repository forks~\cite{cao_what_2022}. These studies typically analyze individual signals in isolation and do not provide a unified view that combines information from package registries and source code repositories.

Several approaches focus on registry metadata. Zahan et al.~\cite{zahan_what_2022} propose security-relevant signals for npm packages with a strong emphasis on maintenance practices, but do not consider repository-level indicators as we do. PyRadar~\cite{kai2024pyradar} addresses the quality of registry metadata by mapping packages to their source repositories, but does not assess broader trust or security implications.

Other tools, such as Socket~\cite{socket} and the OpenSSF Scorecard~\cite{scorecard}, aggregate multiple metadata- and repository-based signals into risk scores or alerts. Empirical evaluations using Scorecard report generally low security scores across projects~\cite{hegewald_evaluating_2025}. Related tooling, such as Heisenberg~\cite{heisenberg}, integrates selected metadata, such as time since release and OpenSSF Scorecard scores, into development workflows.
While such risk scoring approaches provide coarse-grained assessments, they often conflate different signals and emphasize development practices.
They do not support individual decision by developers to add or update a dependency with authenticity and traceability. 

Complementary efforts focus on defining or enforcing integrity and provenance guarantees. SLSA~\cite{slsa} specifies security levels and criteria for supply chain artifacts, particularly around provenance, but does not provide automated, ecosystem-wide assessment. Macaron~\cite{hassanshahi_macaron_2023} operationalizes parts of SLSA by discovering source repositories, checking for provenance, and detecting regressions over time, yet remains focused on compliance with specific standards rather than a broader set of risk indicators. At the ecosystem level, package managers such as PNPM provide policy mechanisms to prevent security regressions, for example, by blocking updates that remove previously available provenance information~\cite{pnpm_trustpolicy}.

Work on reproducible builds proposes strong integrity signals for software artifacts in ecosystems such as Python and Maven~\cite{vltpkg,keshani2024aroma}, but typically treats reproducibility as a standalone property rather than as part of a broader taxonomy.

In contrast, our work focuses on identifying and systematizing individual risk indicators derived jointly from package registry, source code repository, and dependency files, providing a unified and actionable framework rather than a single aggregated score.

\subsection{Dependency Smells and Structural Dependency Issues}

Prior work on dependency smells focuses primarily on maintainability issues rather than supply chain trust or security. 
Existing studies identify problems such as bloated dependencies, missing dependencies, and erroneous version constraints~\cite{jafari_dependency_2022,cao_towards_2023}. Other work aims to reduce attack surface by debloating dependency trees~\cite{Pashakhanloo2022pacjam} or improving dependency resolution reliability~\cite{paulsen2025marco}. While these approaches address important challenges in dependency management, they do not consider trust, authenticity, or traceability of dependencies as we do in this paper.

The concept of smells itself originates from code quality research, where code smells are used as heuristic indicators of technical debt affecting maintenance~\cite{lacerda_code_2020}. Prior work has proposed tools to detect code smells~\cite{cherry_smeagol_2024} and provide recommendations to developers~\cite{de_mello_recommendations_2023}, and more recent studies extend smell-based thinking to build scripts and infrastructure~\cite{tamanna_your_2025}. These approaches demonstrate the usefulness of lightweight, heuristic signals, but focus on internal code and build quality rather than supply chain security.

In contrast, our work adapts the smell abstraction to the software supply chain domain, positioning software supply chain smells as security-oriented signals complementary to maintainability-focused analyses.

\section{Conclusion}

In this paper, we introduced the novel concept of software supply chain smells: they are structural indicators for security issues in modern software supply chains. We presented and evaluated \toolname, an open-source tool for detecting supply chain smells across dependency trees. 
Our interviews with practitioners clearly showed that the proposed smells align with real-world concerns and provide signals practitioners consider meaningful for assessing supply chain risks. 
A quantitative study on packages from Maven and NPM shows the prevalence of smells in both ecosystems; however, with clear differences between them: Traceability and signing issues are prevalent in Maven, while most smells are rare in NPM, partly due to stronger registry-level guarantees.
To conclude, software supply chain smells are pragmatic, grounded, and actionable ways to reason about software supply chain security.

\section*{Acknowledgments}
This work was partially supported by the WASP program funded by Knut and Alice Wallenberg Foundation, and by the Swedish Foundation for Strategic Research (SSF). Some computation was enabled by resources provided by the National Academic Infrastructure for Supercomputing in Sweden (NAISS). 
The authors acknowledge the use of ChatGPT (OpenAI) to improve the language and grammar in the manuscript.

\bibliographystyle{IEEEtran_noAvailable}
\bibliography{bibliography,zoteroreferences}

\end{document}